\documentclass[prd,tightenlines,nofootinbib,prd]{revtex4}
\usepackage{amsmath,amscd,amssymb}
\usepackage{color,epsfig}
\usepackage{xfrac}
\usepackage{latexsym}
\usepackage{mathrsfs}
\usepackage{graphicx}
\usepackage{bbm}      

\newcommand{\imu}{{\rm i}}

\newcommand{\zr}[1]{\mbox{\hspace*{#1em}}}
\newcommand{\ID}{\mbox{{\sf 1}\zr{-0.16}\rule{0.04em}{1.55ex}\zr{0.1}}}
\newcommand{\fract}[2]{{\textstyle\frac{#1}{#2}}}

\begin{document}

\title{Spectral Methods for Coupled Channels with a Mass Gap}

\author{H. Weigel$^{a)}$, M. Quandt$^{b)}$, N. Graham$^{c)}$}

\affiliation{
$^{a)}$Institute for Theoretical Physics, Physics Department, 
Stellenbosch University, Matieland 7602, South Africa\\
$^{b)}$Institute for Theoretical Physics, T\"ubingen University
D-72076 T\"ubingen, Germany\\
$^{c)}$Department of Physics, Middlebury College
Middlebury, VT 05753, USA}

\begin{abstract}
We develop a method to compute the vacuum polarization energy for 
coupled scalar fields with different masses scattering off a 
background potential in one space dimension. 
As an example we consider the vacuum polarization energy
of a kink-like soliton built from two real scalar fields with
different mass parameters.
\end{abstract}

\maketitle

\section{Introduction}

Many non-linear field theories allow for localized field 
configurations that are stable due to their topological properties or, when
they possess non-trivial conserved quantum numbers, because they are 
energetically favored over trivial configurations with the same quantum 
numbers. These configurations are frequently called solitons, even though 
strictly speaking they are merely solitary waves \cite{Ra82}. In specific cases, 
the energetical stabilization mechanism requires the inclusion of quantum
corrections \cite{Farhi:2000ws,Weigel:2010zk}. These so-called vacuum 
polarization energies (VPEs) emerge as the renormalized sums of the shifts of 
the zero point energies for the quantum fluctuations about the soliton. Of 
course, computing the VPE for a soliton configuration is also an interesting
subject even before addressing the stabilization issue. In that calculation 
a major obstacle is the appearance of ultra-violet divergences which are 
characteristic for quantum field theories. Spectral methods \cite{Graham:2009zz} 
have proven efficient at handling these divergences, by computing the VPE from 
scattering data of the potential generated by the soliton and identifying the 
divergences from the corresponding Born series. The latter is then expressed 
as a series of Feynman diagrams that are regularized and renormalized by techniques 
common in perturbative quantum field theory. Still, the full calculation is exact 
and the Born approximation serves only as a mechanism for isolating potentially 
divergent contributions in a tractable form.

Spectral methods, {\it i.e.} the use of quantum scattering calculations, have
a long history in the context of VPEs. The foundation to this approach was formulated 
by Schwinger \cite{Schwinger:1954zz}. Early applications include the renowned work 
published in Ref.~\cite{Dashen:1974cj} and various examples compiled in Ref.~\cite{Ra82}.
In our treatment the Jost function is central, however, the VPE can also be computed
from scattering formulations of the Green's functions \cite{Baacke:1989sb,Graham:2002xq}.
Alternatively, the fluctuation determinant is directly computed (or estimated)
within heat kernel methods \cite{Bordag:1994jz} in conjuction with $\zeta$-function
renormalization \cite{Elizalde:1996zk} (for the present investigation the heat kernel
calculations of Refs.~\cite{AlonsoIzquierdo:2012tw,AlonsoIzquierdo:2003gh} are particularly
relevant), the world line formalism \cite{Gies:2003cv}, derivative expansions \cite{Aitchison:1985pp},
or the Gel'fand-Yaglom method \cite{Gelfand:1959nq}; just to name a few other techniques. 
Many of them have been motivated by research on the famous Casimir effect \cite{Casimir:1948dh}.
For further details (and many more references) on these calculations we refer to pertinent
reviews and text books \cite{Mostepanenko:1997sw,Graham:2009zz}. For the VPE of a single
quantum fluctuation off smooth backgrounds in one space dimension these methods produce 
consistent results. This is mainly due to the fact that merely a single, local Feynman diagram 
must be renormalized. However, in higher dimensional cases, the approximative character 
of some of these approaches causes deviations. Deviations between different approaches have also 
been observed for the VPE in one-dimensional models when un-conventional field 
configurations\footnote{For example, degenerate vacua with different curvatures.} 
are involved as {\it e.g.} for the $\phi^6$ model kink \cite{AlonsoIzquierdo:2011dy}.

Here we extend the spectral method formalism to the multi-threshold case. We begin from $F(k)$, 
the Jost function, or in a multi-channel scattering problem the Jost determinant, where $k$ is 
the momentum of the scattering wave-function. Several analytic properties of $F(k)$ will be 
essential: (i) it must be analytic in the upper half complex $k$-plane; (ii) it must have 
simple zeros for purely imaginary $k=\imu\kappa_i$ representing the bound state solutions; 
and (iii) its phase, {\it i.e.} the physical scattering phase shift, must be anti-symmetric 
for real $k$: $F(-k)=F^\ast(k)$. With mild conditions on the background potential these 
properties are well established even in problems with several \emph{independent} momentum 
like variables $k_1, k_2,\ldots$. That is, a Jost function can be constructed that reflects 
these properties as a function of $k_1$ with $k_2,\ldots$ fixed or, alternatively, as a 
function of $k_2$ with $k_1,\ldots$ fixed \cite{Newton:1982qc,Chadan:1977pq}.

The situation changes drastically when the momentum variables are not
independent as is the case when coupled particles with different masses
scatter off a background potential. As a prototype example, we consider two
particles with masses $m_1$ and $m_2$.  Since the background potential
is static, the energy of the fluctuations is conserved. Using relativistic 
dispersion relations, the momenta are therefore related as
\begin{equation}
k_1^2+m_1^2=k_2^2+m_2^2\,.
\label{eq:dispersion}\end{equation}
We adopt the convention $m_1\le m_2$ so that $k_1$ is real for the physical scattering 
process, and treat $k_1$ as the independent momentum variable. This choice induces a 
square root dependence of $k_2$ on $k_1$ which may lead to additional branch cuts 
in the complex $k_1$ plane. Within the gap, $k_1^2<m_2^2-m_1^2$, the second momentum variable 
becomes complex and, at least in the lower complex half plane, analytic properties
are most likely lost. Within that gap it is also important to ensure that the wave-function 
of the heavier particle decreases exponentially towards spatial infinity. In view of these 
obvious obstacles it is the main objective of the present investigation to establish a 
comprehensive scattering formalism for the VPE in this case, where we generalize 
Eq.~(\ref{eq:dispersion}) to complex momenta such that the above required properties 
of the Jost function $F(k_1)$ are  maintained.

There are additional technical advantages of analytically continuing to complex momenta.
Rotating the momentum integral over the phase shift onto the imaginary axis automatically
includes the bound state contribution to the VPE and they need not be explicitly constructed.
Nevertheless we will do so in certain cases to verify the analytic properties of $F(k_1)$.
Additionally, at threshold, $k_2=0$, the phase shift typically exhibits cusps that are difficult
to handle numerically. This problem is avoided on the imaginary axis, as is the one of 
numerically constructing a phase shift as a continuous function of $k_1$ from the phase of the
determinant of the scattering  matrix without $2\pi$ jumps. Furthermore, as we will discuss 
below, for $m_1\ne m_2$ a consistent formulation of the no-tadpole renormalization condition is 
problematic when the VPE is formulated as an integral over real momenta.

The paper is organized as follows. In Section~\ref{sec:problem} we explain the problem and 
relate $k_1$ and $k_2$ such that the scattering problem is well-defined along the real $k_1$
axis and an analytic continuation to the upper half complex $k_1$ plane is possible. In 
Section~\ref{sec:num1}, we then define the Jost determinant for a spatially symmetric potential
and demonstrate the necessary analytic properties of this determinant by numerical simulations.
In Section~\ref{sec:skewed} we generalize this approach to a potential with mixed reflection 
symmetry similar to parity in the Dirac theory. In Section~\ref{sec:vpe} we formulate the VPE 
in terms of an integral of the Jost determinant along the positive imaginary momentum axis. 
In Section~\ref{sec:bazeia} we apply this spectral method to the soliton model 
with two real scalar fields proposed in Ref.~\cite{Bazeia:1995en} in Section~\ref{sec:bazeia} to 
investigate whether the classical degeneracy of soliton configurations is broken by quantum 
corrections. Our results provide the exact one-loop VPEs, extending previous 
calculations that used the fluctuation spectrum to confirm stability of these 
solutions \cite{Boya:1989dn,Bazeia:1996np,Bazeia:1997zp,Bazeia:1998zv,Veerman:1999kx}.
A short summary is given in Sec.~\ref{sec:concl}. Technical issues are discussed in two
appendixes. In Appendix~\ref{app:cauchy} we verify the analytical structure of the 
Jost determinant by numerically simulating contour integrals and in Appendix~\ref{app:decouple} 
we investigate the VPE for the particular case of two un-coupled particles with a mass gap.

\section{Multi-threshold scattering}
\label{sec:problem}

We consider a field theory in one space  dimension with two scalar fields $\Phi_1$ and 
$\Phi_2$. We write the Lagrangian in terms of the field potential $U(\Phi_1,\Phi_2)$
\begin{equation}
\mathcal{L}=\frac{1}{2}\sum_{i=1}^2\left[\partial_\mu\Phi_i\partial^\mu\Phi_i
-m_i^2\Phi_i^2\right]-U(\Phi_1,\Phi_2)\,,
\label{eq:lag}\end{equation}
and assume that the classical field equations produce a localized static solution
$\Phi_{\rm sol}(x)=(\Phi_{\rm sol,1}(x),\Phi_{\rm sol,2}(x))$. Fluctuating fields 
with frequency $\omega$ are then introduced as 
$\Phi(x,t)=\Phi_{\rm sol}(x)+{\rm e}^{-\imu\omega t}\left(\phi_1(x),\phi_2(x)\right)$. 
For simplicity, the frequency dependence of the fluctuations $\phi_i(x)$ is not made 
explicit. To harmonic order these fluctuations are subject to the wave equation
\begin{equation}
-\frac{d^2}{dx^2}\begin{pmatrix}\phi_1 \cr \phi_2\end{pmatrix}
=\begin{pmatrix} k_1^2 & 0 \cr 0 & k_2^2\end{pmatrix}
\begin{pmatrix}\phi_1 \cr \phi_2\end{pmatrix}
-V(x)\begin{pmatrix}\phi_1 \cr \phi_2\end{pmatrix}\,,
\label{eq:wave1}\end{equation}
where the diagonal momentum matrix arises from the dispersion 
$\omega^2=k_1^2+m_1^2=k_2^2+m_2^2$ and $V(x)$ is the $2\times2$ potential matrix 
that couples the two fluctuating fields via the soliton background
\begin{equation}
V_{ij}(x)=\frac{1}{2}\,\frac{\partial^2 U(\Phi_1,\Phi_2)}
{\partial \Phi_i \partial\Phi_j}\Big|_{\Phi=\Phi_{\rm sol}}\,.
\label{eq:pot1}\end{equation}
At this point we assume $V(x)=V(-x)$, but we will also discuss a different scenario in 
Section \ref{sec:skewed}. Note that for investigating the analytic properties of scattering 
data it is not necessary that the potential arises from a soliton model. 

As mentioned in the introduction, we take $k=k_1$ to be the independent momentum variable.
Then the dependent momentum variable $k_2=k_2(k)$ must have (at least) three important 
properties:
\begin{enumerate}
\item
For any real $k$ within the gap, $k^2<m_2^2-m_1^2$, the dependent momentum 
$k_2$ is imaginary with a \emph{positive} imaginary part, so that $\imu k_2\le0$ 
leads to a localized wave-function in the closed channel.
\item
For any real $k$ outside the gap, $k^2\ge m_2^2-m_1^2$, the dependent momentum 
$k_2$ is also real and $k\to-k$ must imply $k_2\to-k_2$. This will ensure $F(-k)=F^{\ast}(k)$.
\item
If ${\sf Im}(k)\ge0$ then we must also have ${\sf Im}(k_2)\ge0$
such that both momenta are in their respective upper half complex planes.
\end{enumerate}
Properties 1.~and 2.~seem contradictory because the
first one does not allow for a sign change of $k_2$ while the second
one requires it. We will now show that
\begin{equation}
k_2=k_2(k)\equiv k\sqrt{1-\frac{m_2^2-m_1^2}{\left[k+\imu\epsilon\right]^2}}
\qquad {\rm with}\qquad \epsilon\to0^+
\label{eq:k2}\end{equation}
indeed possesses these properties. Property 2.~is obvious because outside the 
gap the real part of the radical is positive and the $\imu\epsilon$ prescription that 
moves the pole into the lower half plane can be ignored. Within the gap, $k^2<m_2^2-m_1^2$,
we expand in $\epsilon$ with careful attention to the sign of its coefficient, yielding
\begin{equation}
k_2(k)=k\sqrt{1-\frac{m_2^2-m_1^2}{k^2}+\imu\,{\rm sign}(k)\epsilon}
= k\imu\,{\rm sign}(k)\sqrt{\frac{m_2^2-m_1^2}{k^2}-1}
=+\imu\sqrt{m_2^2-m_1^2-k^2}\,,
\label{eq:k2a}\end{equation}
satisfying Property 1.~above. Property 3.~is established by introducing 
$k=s+\imu t$ with real $s$ and $t$ in the relation between $k$ and
$k_2$, generalizing Eq.~(\ref{eq:k2a})
\begin{equation}
k_2=(s+\imu t)\sqrt{1-(m_2^2-m_1^2)\frac{s^2-t^2}{(s^2+t^2)^2}
+2\imu\frac{st(m_2^2-m_1^2)}{(s^2+t^2)^2}} \,.
\label{eq:k2b}\end{equation}
We define $X$ and $Y$ such that
\begin{equation}
\sqrt{1-(m_2^2-m_1^2)\frac{s^2-t^2}{(s^2+t^2)^2}
+2\imu\frac{st(m_2^2-m_1^2)}{(s^2+t^2)^2}} = (X+\imu\,{\rm sign}(st)Y)\,.
\end{equation}

The square root halves the phase of any complex number, {\it i.e.} it maps phases 
in $[-\pi,\pi]$ to $[-\pi/2,\pi/2]$, implying $X\ge0$. Also the sign of the imaginary  
part does not change when taking the square root, so that $Y\ge0$ as well. From this 
it follows that ${\sf Im}(k_2)=|s|{\rm sign}(t)Y+tX\ge0$ when $t\ge0$, which is 
the third required property.

With these momentum variables, we are now prepared to parameterize the Jost 
solution as a matrix-valued function. Its columns contain the two functions 
$\phi_1$ and $\phi_2$ subject to specific boundary conditions: Outside the 
gap, the rows refer to the two independent boundary conditions 
of out-going plane waves of the individual particles at positive spatial infinity.
Inside the gap, the second boundary condition parameterizes a localized wave-function
in the closed channel. To this end we apply the differential operator from 
Eq.~(\ref{eq:wave1}) to 
\begin{equation}
F(k,x)=Z(k,x)\begin{pmatrix} {\rm e}^{\imu k x} & 0 \cr
0 & {\rm e}^{\imu k_2(k) x}\end{pmatrix}
\label{eq:Jost1}\end{equation}
and obtain the second order differential equation for the coefficient matrix 
function
\begin{equation}
Z^{\prime\prime}(k,x)=-2Z^\prime(k,x)D(k)+M^2Z(k,x)-Z(k,x)M^2+V(x)Z(k,x)\,,
\label{eq:master}
\end{equation}
where primes denote derivatives with respect to the spatial coordinate
while
\begin{equation}
D(k)=\imu k \begin{pmatrix} 1 & 0 \cr
0 & \sqrt{1-\frac{m_2^2-m_1^2}{\left[k+\imu\epsilon\right]^2}}\end{pmatrix}
\qquad {\rm and }\qquad
M^2=\begin{pmatrix} m_1^2 & 0 \cr 0 & m_2^2\end{pmatrix}
\label{eq:defofDana}
\end{equation}
denote space-independent diagonal matrices. The boundary conditions then translate 
into $\lim_{x\to\infty}Z(k,x)=\ID$ and $\lim_{x\to\infty}Z^\prime(k,x)=0$.

Since Eq.~(\ref{eq:wave1}) is real, the physical scattering solutions are linear 
combinations of $F(k,x)$ and $F^\ast(k,x)=F(-k,x)$. With the assumption that 
$V(-x)=V(x)$ these solutions decouple into symmetric and anti-symmetric channels
\begin{equation}
\Phi_S(k,x)=F(-k,x)+F(k,x)S_S(k)
\qquad {\rm and}\qquad 
\Phi_A(k,x)=F(-k,x)+F(k,x)S_A(k)\,,
\label{eq:smat1}\end{equation}
where $S_S$ and $S_A$ are the scattering matrices that are obtained from the 
boundary conditions $\Phi_S^\prime(k,0)=0$ and $\Phi_A(k,0)=0$ in the respective channels. 
For the VPE we merely require the sum of the eigenphase shifts given by the (logarithm of 
the) determinant of the scattering matrices. Hence we write\footnote{The arguments of the 
determinants on the right hand sides differ from the physical scattering matrix by relative 
flux factors for the off--diagonal elements. In the determinant these factors cancel.}
\begin{equation}
{\rm det}\left[S_S(k)\right]
={\rm det}\left[F^{-1}_S(k)F_S(-k)\right]
\qquad {\rm and}\qquad
{\rm det}\left[S_A(k)\right]
={\rm det}\left[F^{-1}_A(k)F_A(-k)\right]
\label{eq:smat2}\end{equation}
with the Jost matrices
\begin{equation}
F_S(k)=\lim_{x\to0}\left[Z^\prime(k,x)D^{-1}(k)+Z(k,x)\right]
\qquad {\rm and}\qquad
F_A(k)=\lim_{x\to0}Z(k,x)\,.
\label{eq:defjostsym}
\end{equation}
In the symmetric channel, alternative definitions of the Jost matrix $F_S(k)$ with 
different factors of $D(k)$ lead to the same determinant of $S_S$, because $D^\ast(k)=D(-k)$.
This ambiguity is removed by demanding that $F_S(k)$ be the unit matrix for 
$V\equiv0$, which is equivalent to requiring $\lim_{k\to\infty}F_S(k)=\ID$. The factor 
of $D(k)$ as in Eq.~(\ref{eq:defjostsym}) adds multiples of $\pi$ to to the phase shift 
when $k^2\le m_2^2-m_1^2$. In Appendix \ref{app:decouple} we show that 
these multiples of $\pi$ produce the correct VPE for the specific case that the two 
particles are uncoupled, {\it i.e.}~when the potential matrix $V$ is diagonal.

\section{Numerical Experiments for Symmetric Potential Matrix}
\label{sec:num1}

In this Section use a minimal example to show that the determinants of the matrices 
defined in Eq.~(\ref{eq:defjostsym}), ${\rm det}\left[F_S(k)\right]$ 
and ${\rm det}\left[F_A(k)\right]$, have the following properties:
\begin{itemize}
\item[$\bullet$] 
for real $k$, their real and imaginary parts are even and odd functions, respectively.
This implies that their phases are odd (more precisely, the momentum derivatives
of their phases are even),
\item[$\bullet$]
they have no singularities or branch cuts for ${\sf Im}(k)\ge0$,
\item[$\bullet$]
for a particular form of Levinson's theorem the phases at $k=0$ can be used to count
the number of bound states, and
\item[$\bullet$]
they have zeros along the positive imaginary axis $k=\imu\kappa_i$ corresponding to 
bound state energies $\omega_i^2=m_1^2-\kappa_i^2$. (For strong enough attraction,
$\omega_i^2$ may be negative, indicating instability. For the models to be investigated 
in Section \ref{sec:bazeia}, this is not the case.) For decoupled channels, $V_{12}=0$, 
bound states of the heavier particle yield zeros at imaginary $k_2$. If these
bound states have energy eigenvalues $m_1^2<\omega_i^2<m_2^2$, these zeros are on the 
real $k$-axis. When $V_{12}\ne0$, the heavier particle can decay into the lighter one 
and these bound states become \emph{Feshbach resonances} \cite{Feshbach:1958nx}.
\end{itemize}
While the first two items serve as tests for the properties established in the 
previous Section, the latter two provide important information about the zeros of 
the Jost determinant. They are crucial when it comes to compute integrals involving
the scattering phase shift, {\it i.e.} the logarithm of the Jost determinant, by means
of Cauchy integral rules.

We will confirm these properties by numerical experiments using a specific, but 
generic potential matrix,
\begin{equation}
V(x)=V_0\,{\rm exp}\left(-\frac{x^2}{w^2}\right)
\qquad {\rm with}\qquad 
V_0=\begin{pmatrix}v_{11} & v_{12} \cr v_{12} & v_{22}\end{pmatrix}\,,
\label{eq:testpot1}\end{equation}
which we can easily tune to have an arbitrary number of bound states
by changing the width $w$ and the entries of the constant symmetric matrix $V_0$.
In the numerical simulations we vary the parameter $\epsilon$ from Eq.~(\ref{eq:k2})
in the range  $\epsilon\in[10^{-40},10^{-20}]$.

\begin{figure}
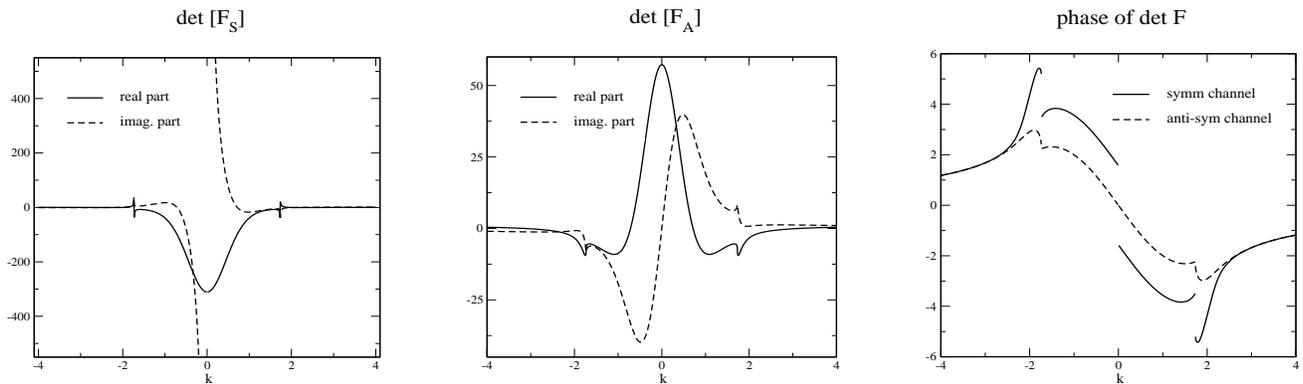

\centerline{
\epsfig{file=detfs1.eps,width=5.0cm,height=5cm}\hspace{1.0cm}
\epsfig{file=detfa1.eps,width=5.0cm,height=5cm}\hspace{1.0cm}
\epsfig{file=detph1.eps,width=5.0cm,height=5cm}}
\caption{\label{fig:jost1} The left and middle panels show the Jost determinants 
in the symmetric and anti-symmetric channels for the potential, Eq.~(\ref{eq:testpot1}) 
with $v_{11}=4.0,\,v_{22}=1,\, v_{12}=0.5$ and $w=2.0$. The mass parameters 
are $m_1=1$ and $m_2=2$. The right panel shows their phases for the same set 
of parameters.} 
\end{figure}

\begin{figure}
\centerline{
\epsfig{file=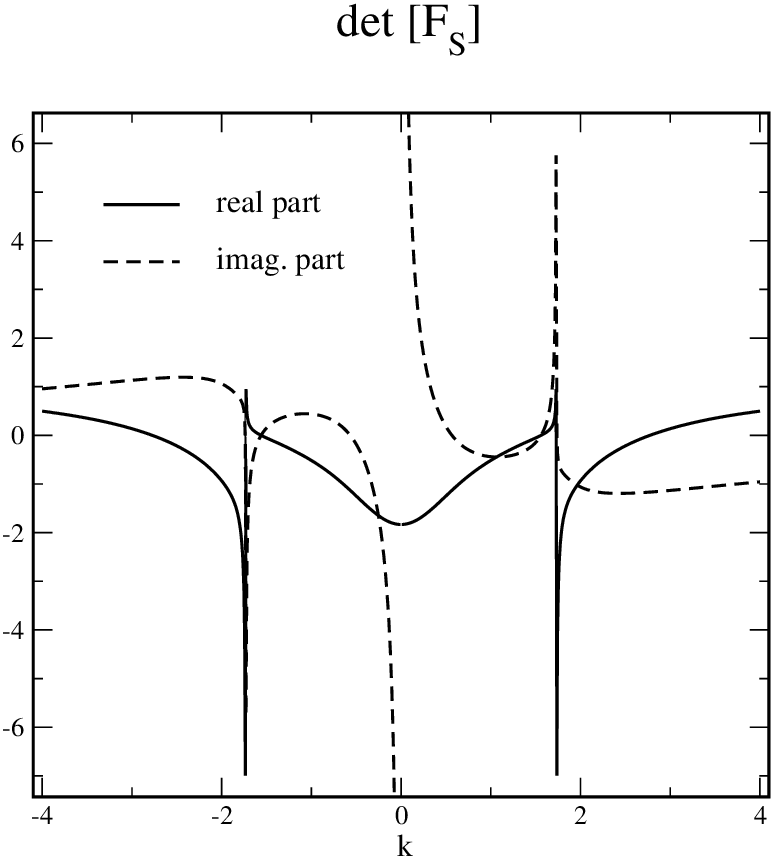,width=5.0cm,height=5cm}\hspace{1.0cm}
\epsfig{file=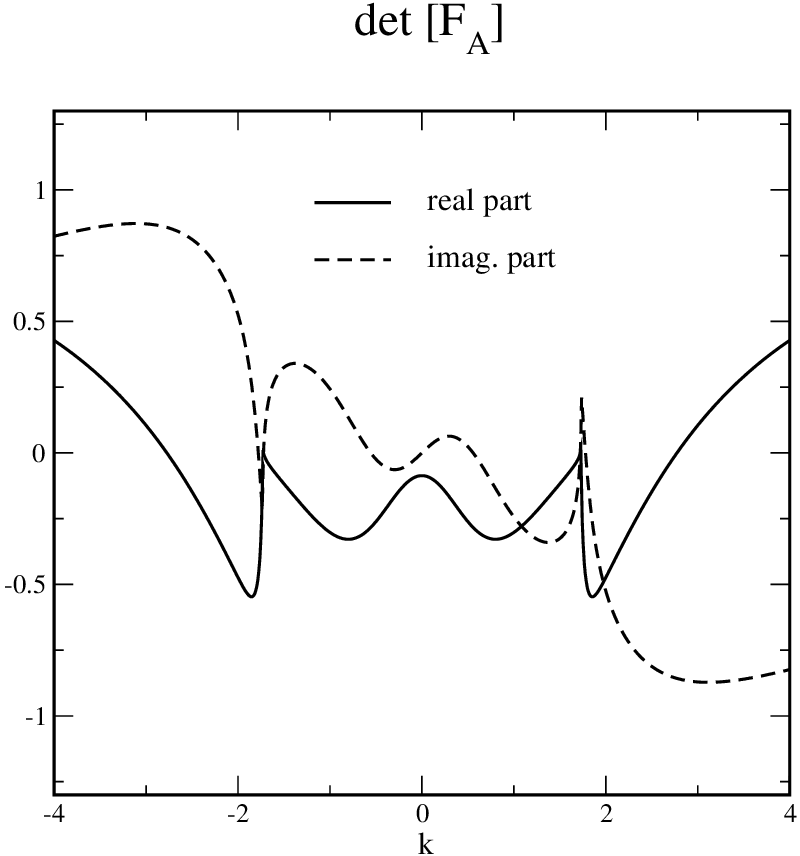,width=5.0cm,height=5cm}\hspace{1.0cm}
\epsfig{file=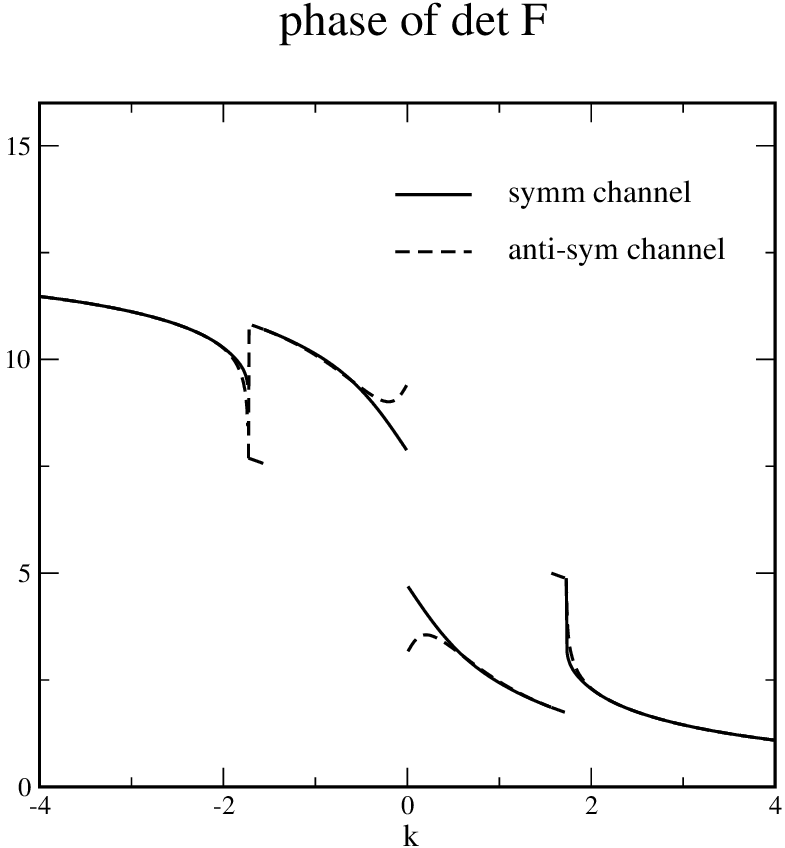,width=5.0cm,height=5cm}}
\caption{\label{fig:jost2} Same as figure \ref{fig:jost1} with $V_0\to-V_0$.}
\end{figure}

\medskip\noindent
In Figs.~\ref{fig:jost1} and \ref{fig:jost2} we show the Jost determinants and 
their phases for repulsive and attractive potentials, respectively. The phases are
computed as
$$
\delta(k)=- {\sf Im}\,{\rm log}\,{\rm det}F(k)
= {\sf Im}\,{\rm log}\,{\rm det}F(-k)
\,.
$$
The second equality gives the standard definition of the scattering phase 
shift~\cite{Newton:1982qc} and is valid only if ${\rm det}F^\ast(k)={\rm det}F(-k)$.
The proper branch of the logarithm is determined by eliminating $2\pi$ jumps between 
neighboring points and imposing $\lim_{k\to\infty}\delta(k)=0$.

The real and imaginary parts of the Jost determinants are clearly even and 
odd functions of the single momentum variable $k$, respectively. Discontinuities or 
even singularities may occur at threshold $k^2=m_2^2-m_1^2$ and at $k=0$, which 
combine with jumps in the phases when bound states exist as required by 
Levinson's theorem.  For repulsive potentials, this theorem implies \cite{Barton:1984py}
$\lim_{k\to0^+}\delta_S(k)=-\frac{\pi}{2}$. Hence the only phase without a jump is for a 
repulsive potential in the anti-symmetric channel. Yet, that phase still exhibits threshold 
\emph{cusps}. For the attractive potential we also observe rapid changes
slightly below threshold. They reflect Feshbach resonances, also called 
bound states in the continuum \cite{Newton:1982qc}. In total, the Jost determinants defined 
in Eq.~(\ref{eq:defjostsym}) combined with Eq.~(\ref{eq:k2}) have exactly the expected 
features describing physical scattering for real $k$. We have checked these results for
other values of $V_0$ and $w$ as well.

A particularly interesting case is that of a system with interactions only in the 
heavy particle channel. The phase shift is treated as a function of the lighter particle's 
momentum and therefore must vanish inside the gap, or at least be a (piecewise constant) 
multiple of $\pi$. If it did not, the VPE would not be the simple sum of the two 
particles' VPEs when the potential matrix is diagonal.
\begin{figure}
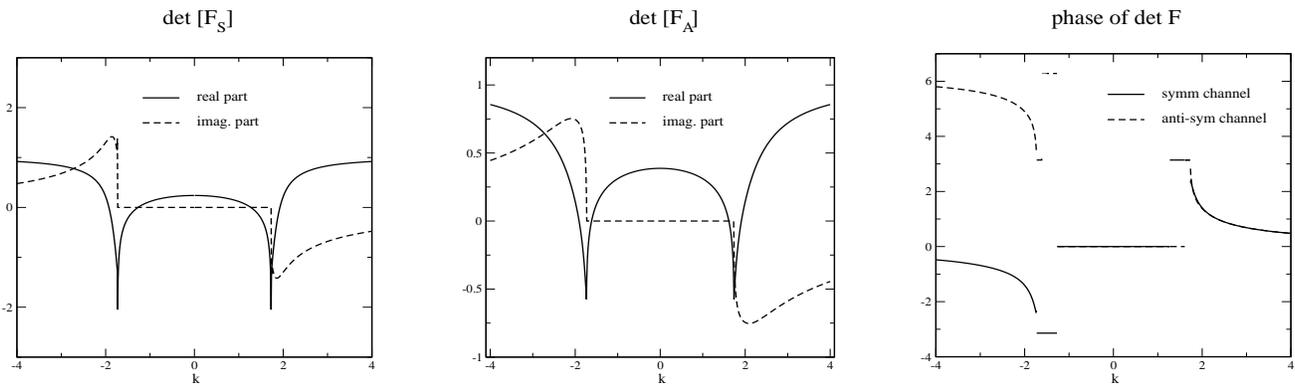

\centerline{
\epsfig{file=detfs3.eps,width=5.0cm,height=5cm}\hspace{1.0cm}
\epsfig{file=detfa3.eps,width=5.0cm,height=5cm}\hspace{1.0cm}
\epsfig{file=detph3.eps,width=5.0cm,height=5cm}}
\caption{\label{fig:jost_dec} Real and imaginary parts and phases of the 
Jost determinants, Eq.~(\ref{eq:defjostsym}) for $v_{11}=v_{12}=0$,
$v_{22}=-2$ and $w=2$.}
\end{figure}
This property is verified in Figure \ref{fig:jost_dec}.  This result relies crucially on 
including the kinematic factors exactly as in Eq.~(\ref{eq:defjostsym}). Also, the jumps 
by $\pi$ within the gap are significant for the VPE, as will be discussed in 
Appendix \ref{app:decouple}. 

For the parameters of Figure \ref{fig:jost_dec}, a close inspection of the symmetric 
channel reveals that the phase shift jumps from $0$ to $\pi$ at $k=1.270$ and from 
$\pi$ to $\frac{\pi}{2}$ at threshold. These jumps hamper the continuous construction
of phase shifts. In the anti-symmetric channel a jump occurs at $k=1.615$, while this phase 
shift is continuous at threshold. Even though the imaginary parts of the Jost determinants are 
zero within the gap, the jumps at $k=1.270$ and $k=1.615$ result from sign changes of the 
real parts as can also be observed from Figure \ref{fig:jost_dec}. When the off-diagonal 
elements of $V_0$ are switched on, ${\sf Im}\,{\rm det}F\ne0$ and ${\rm det}F$ has no zeros 
along the real axis. At the same time, the bound states from the uncoupled situation turn
into (sharp) Feshbach resonances corresponding to zeros in the lower half of the complex 
momentum plane. We will comment further on this point at the end of Appendix \ref{app:cauchy}.

This discussion also reveals another interesting property of the phase shift related to
Levinson's theorem. For single-channel scattering, this theorem states that the phase shift 
at zero momentum is $\pi$ times the number of bound states in the anti-symmetric channel, 
However, for the symmetric channel, the ratio between the phase shift at zero momentum and 
$\pi$ equals the number of bound states minus $\frac{1}{2}$ \cite{Barton:1984py}.
For a two-channel problem one might therefore expect the phase shift at zero momentum to be
$\pi$ times the number of bound states minus 1.  The right panels of Figures \ref{fig:jost1}, 
\ref{fig:jost2} and \ref{fig:jost_dec} show that this is not the case. Rather, the discontinuities 
that result from Levinson's theorem appear at the respective thresholds. This is evident for 
the uncoupled system, and persists when the off-diagonal elements of the potential matrix are 
switched on, though the jump is slightly smoothed out in that case.

To identify genuine bound states as well as Feshbach resonances independently from the Jost 
function calculation we diagonalize the Hamiltonian
\begin{equation}
H=-\ID\frac{d^2}{dx^2}+\begin{pmatrix}m_1^2 & 0 \cr 0 & m_2^2\end{pmatrix}+V(x)
\label{eq:Ham}\end{equation}
associated with the field Equation (\ref{eq:wave1}) in a discretized basis.
The $2N$ basis wave-functions in the symmetric channel are 
\begin{equation}
\Phi^{(0)}_n(x)=\frac{1}{\sqrt{2L}}\begin{pmatrix}{\rm cos}(p_nx) \cr 0\end{pmatrix}
\qquad {\rm and}\qquad
\Phi^{(0)}_{n+N}(x)=\frac{1}{\sqrt{2L}}\begin{pmatrix}0 \cr {\rm cos}(p_nx)\end{pmatrix}\,,
\label{eq:basis1}\end{equation} 
with $x\in[0,L]$, which is sufficient since the potential is symmetric. The 
discretized momenta $p_n$ are determined such that $\Phi^{(0)}(L)=0$. 
Replacing ${\rm cos}(px)\,\rightarrow\,{\rm sin}(qx)$ gives the basis wave-functions in the 
anti-symmetric channel, where we must also redefine the discretized momenta 
$p_n\to q_n$ in order to maintain $\Phi^{(0)}(L)=0$. In general the eigenvalues $\omega^2$ of $H$ 
depend on $L$. However, this is not the case for bound states when $L$ is sufficiently large, 
because their wave-functions vanish at spatial infinity. Similarly the Feshbach resonances are 
associated with those eigenvalues in the continuum that do not vary when $L$ is altered.
For the parameters of Figure \ref{fig:jost2} we observe bound state eigenvalues
at $\omega^2 = -2.128$ and $\omega^2 = 0.603$ in the symmetric channel. They correspond to 
$\kappa=1.768$ and $\kappa = 0.630$. This agrees with Levinson's theorem, since
the phase shift at zero momentum equals $\frac{3\pi}{2}$. In the anti-symmetric 
channel only a single eigenvalue at $\omega^2= -0.513$ with $\kappa=1.230$ occurs.
Accordingly, the phase shift equals $\pi$ at zero momentum. In the symmetric channel, 
a Feshbach resonance appears with the eigenvalue $\omega^2 = 3.435$, {\it i.e.} $k=1.560$,
causing the corresponding phase shift to be $\frac{\pi}{2}$ at threshold. The Feshbach 
resonance in the anti-symmetric channel is just below threshold. It has the eigenvalue 
$\omega^2 = 3.968$, {\it i.e.} $k=1.723$ and the phase shift jumps by $\pi$ at threshold. 
Increasing the potential strength slowly moves this eigenvalue deeper into the gap, 
{\it e.g.} $v_{22}=1.2$ yields the eigenvalue $\omega^2 = 3.922$, while the resonance 
disappears for $v_{22}=0.6$ and the phase shift becomes continuous at threshold, though 
the typical cusps remains. We conclude that the Jost determinant as defined in 
Eq.~(\ref{eq:defjostsym}) together with the definition of the dependent momentum variable, 
Eq. (\ref{eq:k2}), possess all required properties for real momenta. 

Finally, it is interesting to compute the $\omega^2$ eigenvalues for the parameters of 
Figure \ref{fig:jost_dec}, {\it i.e.} the case when the lighter particle does not interact. 
We have observed bound states below $m_2^2$ at $\omega^2=2.613$ and $\omega^2=3.608$ in the 
symmetric and anti-symmetric channels, respectively. These bound state energies 
correspond to momenta $k_2$ on the positive imaginary axis that relate to real $k$
when $\omega^2>m_1^2$. Hence at these values of $k$ the phase shifts jump by $\pi$, 
as shown in the right panel of Figure \ref{fig:jost_dec}. 

\begin{figure}
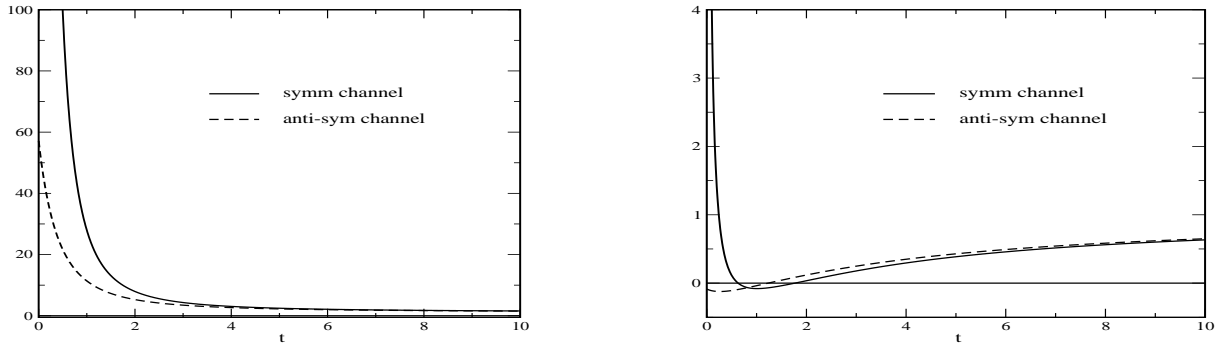

\centerline{
\epsfig{file=detFt1.eps,width=7.0cm,height=5cm}\hspace{2cm}
\epsfig{file=detFt2.eps,width=7.0cm,height=5cm}}
\caption{\label{fig:detFt}The real Jost determinant for $k=\imu t$. The left and right 
panels corresponds to the parameters of Figures \ref{fig:jost1} and \ref{fig:jost2},
respectively. Note the different scales for the vertical axes.}
\end{figure}

When the coupling to the lighter particle is switched on again, these bound states 
turn into Feshbach resonances because the heavier particle may decay into the lighter 
one. Then a key additional property of the Jost determinant is that its only
zeros (simple zeros in single channel scattering) are at purely imaginary bound 
state momenta $k = k_1 =\imu\kappa_i$. Therefore we consider the Jost determinant for 
real~$t$ defined via $k=\imu t$, so that the differential equation, Eq.~(\ref{eq:wave1}),
is purely real (the $\imu\epsilon$ prescription can be omitted for $t>0$) and
so is the Jost determinant. In Figure \ref{fig:detFt} we show the 
corresponding numerical results.
The positions of the zeros of ${\rm det}F$ clearly agree with the above listed
values for $\kappa_i$. In the symmetric channel the determinant diverges as $t\to0$
because of $D^{-1}$ in Eq.~(\ref{eq:defjostsym}). This is actually the origin for the 
$\frac{\pi}{2}$ in Levinson's theorem. This divergence is not a serious problem here 
because the VPE is only sensitive to $t\ge m_1$. At large $t$ the determinant approaches 
unity, though this asymptotic value is assumed only slowly. Without any bound states the 
Jost determinant is positive along the positive imaginary axis.

In appendix \ref{app:cauchy} we present further numerical analysis of the Jost determinant
using Cauchy integrals.

\section{Skewed Parity}
\label{sec:skewed}

So far we have studied the case of a completely symmetric potential matrix. For the model of 
Ref.~\cite{Bazeia:1995en}, we will consider potential matrices symmetric under skewed parity,
\begin{equation}
V(-x)=\begin{pmatrix} 1 & 0 \cr 0 & -1\end{pmatrix}
V(x)\begin{pmatrix} 1 & 0 \cr 0 & -1\end{pmatrix}\,.
\label{eq:skewedV}\end{equation}
That is, diagonal and off--diagonal elements of the potential
matrix are even and odd functions of the coordinate, respectively.
Similar to the Dirac theory we introduce a parity operator
\begin{equation}
\hat{P}\Phi(x)=\hat{P}\begin{pmatrix}\phi_1(x) \cr \phi_2(x)\end{pmatrix}
=\begin{pmatrix}\phi_1(-x) \cr -\phi_2(-x)\end{pmatrix}
\label{eq:skewedP}\end{equation}
that has eigenvalues $\lambda_p=\pm1$. It is then convenient
to define the projectors
\begin{equation}
P_{+}=\begin{pmatrix}1 & 0 \cr 0 & 0\end{pmatrix}
\qquad {\rm and}\qquad
P_{-}=\begin{pmatrix}0 & 0 \cr 0 & 1\end{pmatrix}\,,
\label{eq:parity}\end{equation}
so that $\phi_1(x)=P_{+}\Phi(x)$ and $\phi_2(x)=P_{-}\Phi(x)$.
In this notation the channel with $\lambda_p=1$ has
$$
P_{+}\partial_x\Phi_{+}(0)=0
\qquad {\rm and}\qquad
P_{-}\Phi_{+}(0)=0\,,
$$
while the $\lambda_p=-1$  channel has
$$
P_{-}\partial_x\Phi_{-}(0)=0
\qquad {\rm and}\qquad
P_{+}\Phi_{-}(0)=0\,.
$$
The Jost matrices for the two parity channels can also be obtained 
from the solution of Eq.~(\ref{eq:master}). Suitable projection yields
\begin{equation}
F_{\pm}(k)=\left[P_{\pm}F_S(k)D_{\mp}(k)+P_{\mp}F_A(k)D_{\pm}^{-1}(k)\right]\,.
\label{eq:skjost}
\end{equation}
The factor matrices
$$
D_{+}(k)=\begin{pmatrix}\imu k & 0 \cr 0 & 1\end{pmatrix}
\qquad {\rm and}\qquad
D_{-}(k)=\begin{pmatrix}1 & 0 \cr 0 & \imu k_2\end{pmatrix}\,,
$$
with $k_2$ being again related to $k$ via Eq.~(\ref{eq:k2}),
ensure that $\lim_{|k|\to\infty}F_{\pm}(k)=\ID$. Note that any right multiplication
of $F(k,x)$ in Eq.~(\ref{eq:Jost1}) by a constant matrix is a solution to the field 
Equation~(\ref{eq:master}) and translates into a right multiplication of $Z(k,x)$
when this constant matrix is diagonal.\footnote{Various columns/rows of $F_{\pm}$ may
have different dimensions, but ${\rm det}(F_{\pm})$ has a well defined dimension, 
as does the scattering matrix $S_{\pm}=F^\ast_{\pm}F^{-1}_{\pm}$.}

In the positive parity channel the bound state energies, and eventually the 
position of the Feshbach resonances, are obtained by diagonalizing the Hamiltonian, 
Eq.~(\ref{eq:Ham}) with respect to the $2N$ basis states
\begin{equation}
\Phi^{(0)}_n(x)=\frac{1}{\sqrt{2L}}\begin{pmatrix}{\rm cos}(p_nx) \cr 0\end{pmatrix}
\qquad {\rm and}\qquad
\Phi^{(0)}_{n+N}(x)=\frac{1}{\sqrt{2L}}\begin{pmatrix}0 \cr {\rm sin}(q_nx)\end{pmatrix}\,.
\label{eq:basis2}\end{equation}
Again the discretized momenta $p_n$ and $q_n$ are obtained from $\Phi^{(0)}(L)=0$.
The basis states in the negative parity channel are constructed by swapping upper and
lower components.

Using the trial potential matrix
$$
V(x)=V_1(x)\,{\rm exp}\left(-\frac{x^2}{w^2}\right)
\qquad {\rm with}\qquad 
V_1(x)=\begin{pmatrix}v_{11} & v_{12}\,x \cr v_{12}\,x & v_{22}\end{pmatrix}
$$
we have repeated the numerical experiments from Section \ref{sec:num1}
and appendix \ref{app:cauchy}
and established that the determinants defined by the matrices of 
Eq.~(\ref{eq:skjost}) have the standard Jost function properties
discussed in Section \ref{sec:problem}.

\section{Formulation of Vacuum Polarization Energy}
\label{sec:vpe}

Formally the VPE is the sum of the shifts of zero-point energies due to the 
interaction with the potential $V(x)$. It can be decomposed as \cite{Graham:1999pp}
\begin{equation}
E_{\rm vac}=\frac{1}{2}\sum_{\omega_i^2\le m_1^2}\left[\omega_i-m_1\right]
+\int_0^\infty \frac{dk}{2\pi}\,\left[\omega(k)-m_1\right]
\frac{d}{dk}\sum_s \delta_s(k)\Big|_{\rm ren.}\,.
\label{eq:evac1}\end{equation}
where the discrete sum is over the bound state energies (excluding Feshbach resonances)
and $k$ labels the continuum scattering states with energy $\omega(k)$ such that $k=0$
denotes the scattering threshold. According to the Krein formula \cite{Faulkner:1977aa} 
the derivative of the total phase shift, 
\begin{equation}
\delta_s(k)=\frac{1}{2\imu}{\rm ln}\,{\rm det}S_s(k)
=\frac{1}{2\imu}\left[{\rm ln}\,{\rm det}F_s(-k)-{\rm ln}\,{\rm det}F_s(k)\right]
\label{eq:phase1}\end{equation}
in channel $s$ gives the change in the density of scattering states. These channels can either 
be the symmetric and anti-symmetric channels discussed in Sections \ref{sec:problem} and 
\ref{sec:num1}, or the parity modes from the previous Section. We note that Eq.~(\ref{eq:evac1}) 
can also be derived from the energy-momentum tensor in field theory~\cite{Graham:2002xq}.
Finally the subscript ``ren.'' indicates that, as it stands, the integral diverges and requires
regularization and renormalization. Working in one space dimension, the Born approximation
\begin{equation}
\delta^{(1)}(k)=-\frac{1}{k}\int_0^\infty dx\,V_{11}(x)
-\frac{1}{\sqrt{k^2-m_2^2+m_1^2}}\int_0^\infty dx\,V_{22}(x)
\label{eq:born1}\end{equation}
is typically subtracted under the integral to impose the no-tadpole renormalization condition, 
which removes all terms linear in $V(x)$ from the VPE. While the pole at $k=0$ is canceled by 
the factor $\omega(k)-m_1\,\rightarrow\,\fract{k^2}{2m_1}$ as $k\to0$, the singularity at 
threshold $k=\sqrt{m_2^2-m_1^2}$ precludes a direct application of this prescription when 
different mass parameters are involved. Since 
\begin{equation}
\Delta^{(1)}(k)=-\frac{1}{k}\int_0^\infty dx\,V_{11}(x)
-\frac{1}{k}\int_0^\infty dx\,V_{22}(x)
\label{eq:born2}\end{equation}
has the same large $k$ behavior as $\delta^{(1)}(k)$, it serves well as a helper function 
such that 
\begin{equation}
\widetilde{E}_{\rm vac}=\frac{1}{2}\sum_{\omega_i^2\le m_1^2}\left[\omega_i-m_1\right]
+\int_0^\infty \frac{dk}{2\pi}\,\left[\omega(k)-m_1\right]
\frac{d}{dk}\left[\left(\sum_s \delta_s(k)\right)-\Delta^{(1)}(k)\right]
\label{eq:evac2}\end{equation}
is finite. In contrast to $\delta^{(1)}(k)$ which has a square root branch cut,
$\Delta^{(1)}(k)$ is analytic up the pole at $k=0$ which is removed by the factor
$\omega(k)-m_1$.

Although $\widetilde{E}_{\rm vac}$ is finite, it does not obey the no-tadpole 
renormalization condition. By analytic continuation we will next obtain an integral 
expression for the VPE that allows us to replace $\Delta^{(1)}$ by $\delta^{(1)}$ and
also avoids integrating over the various cusps (or even jumps) of the phase shift within 
the gap. In a first step we use Eq.~(\ref{eq:phase1}), replace $\delta_s(k)$ by the odd 
function ${\rm ln}\,{\rm det}F_s(k)$ and express the VPE as a single integral along the 
whole real axis, 
$$
\widetilde{E}_{\rm vac}=\frac{1}{2}\sum_{\omega_i^2\le m_1^2}\left[\omega_i-m_1\right]
-\int_{-\infty}^\infty \frac{dk}{4\pi\imu}\,\left[\omega(k)-m_1\right]\frac{d}{dk}
\left[\left(\sum_s {\rm ln}\,{\rm det}F_s(k)\right) -\Delta^{(1)}(k)\right]\,.
$$
As numerically verified in Appendix \ref{app:cauchy},
$\frac{d}{dk}{\rm ln}\,{\rm det}F_s(k)$
has first order poles with unit residue exactly at those imaginary momenta that 
represent the bound state energies $\omega_i$. When closing the contour in the upper half
plane, these poles compensate the explicit contributions from the bound states, 
{\it i.e.} the first term in the above equation. The only contribution then comes from 
enclosing the branch cut along the imaginary axis ($t\in\mathbb{R}$)
$$
\sqrt{m_1^2+(\imu t+\epsilon)^2}-\sqrt{m_1^2+(\imu t-\epsilon)^2}=2\imu\sqrt{t^2-m_1^2}
\qquad {\rm for}\quad t\ge m_1\,,
$$
yielding
$$
\widetilde{E}_{\rm vac}=-\int_{m_1}^\infty \frac{dt}{2\pi}\,\sqrt{t^2-m_1^2}\,
\frac{d}{dt}\left[\left(\sum_s {\rm ln}\,{\rm det}F_s(\imu t)\right) -\Delta^{(1)}(\imu t)\right]\,.
$$
Finally we integrate by parts and obtain
\[
\widetilde{E}_{\rm vac}=\int_{m_1}^{\infty} \frac{dt}{2\pi} 
\frac{t}{\sqrt{t^2-m_1^2}}\,\left[\nu(t)-\Delta^{(1)}(t)\right]
\]
where 
\begin{equation}
\nu(t) \equiv {\rm ln}\,{\rm det}\left[F_{+}(\imu t) \,F_{-}(\imu t)\right]
\label{eq:defnu}
\end{equation}
denotes the summed exponentials of the Jost determinants. Without difficulty we
can now restore the original no-tadpole scheme, 
\begin{equation}
E_{\rm vac} \equiv 
\int_{m_1}^{\infty} \frac{dt}{2\pi} 
\frac{t}{\sqrt{t^2-m_1^2}}\,\left[\nu(t)-\nu^{(1)}(t)\right]\,,
\label{eq:Evac}
\end{equation}
where we have replaced $\Delta^{(1)}(\imu t)$ with
\begin{equation}
\nu^{(1)}(t)=\int_{0}^\infty dx\, \left[\frac{V_{11}(x)}{t}
+\frac{V_{22}(x)}{\sqrt{t^2+m_2^2-m_1^2}}\right]\,.
\label{eq:Born1}\end{equation}
It is obvious that this subtraction fully removes the $\mathcal{O}(V)$ contribution to the VPE. 
Up to a factor of $\imu$ it is the continuation of the Born approximation, Eq.~(\ref{eq:born1}), 
to the imaginary axis. Formally Eqs.~(\ref{eq:Evac})-(\ref{eq:Born1}) are 
standard \cite{Graham:2009zz}, except for the second denominator in the Born approximation. 
It clearly reflects the infra-red singularity at $k^2=m_2^2-m_1^2$ above. Fortunately, that 
is not an issue when we integrate along the imaginary axis.

Without off-diagonal interactions $\nu$ and $\nu^{(1)}$ are the sums of the Jost functions for 
the two particles with the imaginary momentum variables $t$ and $\sqrt{t^2+m_2^2-m_1^2}$, 
respectively. Thus it is obvious that Eq.~(\ref{eq:Evac}) is just the sum of the individual 
contributions, as it should be. This is also the case for the real momentum formulation, 
Eq.~(\ref{eq:evac2}). However, the additivity is less obvious on the real axis because it 
requires careful consideration of the phase shift within the gap, as we will discuss in 
Appendix \ref{app:decouple}.

\section{Bazeia Model}
\label{sec:bazeia}

The Bazeia model \cite{Bazeia:1995en} generalizes the $\phi^4$ kink model in one space 
and one time dimensions by introducing a second scalar field $\chi$. Since the model 
admits various soliton solutions that are classically degenerate it is a prime candidate
to apply the above developed method to compute the VPE and investigate whether this 
degeneracy is lifted by quantum corrections. In  this Section, we adopt the notation for 
the fields from the original  paper \cite{Bazeia:1995en},
$(\Phi_1,\Phi_2)\,\longrightarrow\,(\phi,\chi)$. The mass terms are included in the field 
potential, $U(\phi,\chi)$, in contrast to Eq.~(\ref{eq:lag}).

After appropriate rescaling of the fields and the coordinates the Lagrangian
\begin{equation}
\mathcal{L}=\frac{1}{2}\left[\partial_\mu \phi\partial^\mu \phi
+\partial_\mu \chi\partial^\mu \chi\right]-U(\phi,\chi)\,,
\qquad {\rm with}\qquad
U(\phi,\chi)=\frac{1}{2}\left[\phi^2-1+\frac{\mu}{2}\chi^2\right]^2
+\frac{\mu^2}{2}\phi^2\chi^2
\label{eq:fpot}
\end{equation}
is characterized by the single parameter $\mu$. The vacuum configuration is 
$\phi_{\rm vac}=\pm1$ and $\chi_{\rm vac}=0$ so that $m_\chi=\mu$ and $m_\phi=2$. 
Subtracting the mass terms from $\partial^2 U$ yields the fluctuation potential matrix
\begin{equation}
V \equiv 
\partial^2 U - \begin{pmatrix}\mu^2 & 0 \\[2mm] 0 & 4\end{pmatrix}
=\begin{pmatrix}
\mu(1+\mu)\left(\phi^2-1\right)+\frac{3}{2}\mu^2\chi^2 &
2\mu(1+\mu)\chi\phi \\[2mm] 2\mu(1+\mu)\chi\phi &
6\phi^2-6+\mu(\mu+1)\chi^2
\end{pmatrix}\,.
\label{eq:vpot}
\end{equation}
Later we will be interested solely in the case $\mu\le2$.
Then $\chi$ is the lighter particle and hence plays the role of $\Phi_1$.

There is always the pure kink soliton
\begin{equation}
\phi^{(1)}={\rm tanh}(x)
\qquad {\rm and}\qquad 
\chi^{(1)}=0\,.
\label{eq:sol1}
\end{equation}
In addition, for $\mu\le1$ a second solution is 
\begin{equation}
\phi^{(2)}={\rm tanh}(\mu x)
\qquad {\rm and}\qquad 
\chi^{(2)}=\frac{\sqrt{2\left(1/\mu-1\right)}}{{\rm cosh}(\mu x)}\,.
\label{eq:sol2}
\end{equation}
The construction of these solitons is described in Ref.~\cite{Bazeia:1995en}.
For $\mu=1$ the two solutions are identical. They are actually degenerate in 
classical mass, $E_{\rm cl}=\frac{4}{3}$ since the model is defined such that 
$E_{\rm cl}$ can be uniquely determined from the asymptotic values of the fields.  
The specific cases of the fluctuation potential Eq.~(\ref{eq:vpot}) for the two
soliton solutions (\ref{eq:sol1}) and (\ref{eq:sol2}) are
\begin{align}
V^{(1)}&=\frac{-1}{{\rm cosh}^2(x)}
\begin{pmatrix}
\mu(1+\mu) & 0 \\[2mm] 0 & 6\end{pmatrix}\,,
\label{eq:vpot1a}\\[3mm]
V^{(2)}&=\frac{1}{{\rm cosh}^2(x)}
\begin{pmatrix}
2\mu-4\mu^2\qquad &
2(1+\mu)\sqrt{2\mu(1-\mu)}\,{\rm sinh}(\mu x)\\[2mm]
2(1+\mu)\sqrt{2\mu(1-\mu)}\,{\rm sinh}(\mu x) & -4-2\mu^2
\end{pmatrix}\,.
\label{eq:vpot1b} 
\end{align}
Since $\phi$ is the heavier particle, the translational zero mode wave-function is odd in 
the upper component and even in the lower one. Hence this zero mode lives in the negative 
parity channel. However, we also observe a zero mode in the positive parity channel. This 
second zero mode can be used to construct a family of soliton solutions parameterized by 
$|\chi(0)|\le\sqrt{\fract{2}{\mu}}$. This family has already been found in the context 
of supersymmetric domain wall models \cite{Shifman:1997wg}. Except for the particular 
cases $\mu=\fract{1}{2}$ and $\mu=2$, which lead to the solitons discussed in 
Eqs.~(\ref{eq:sol3}) and ~(\ref{eq:sol4}) below, these solitons can only be formulated 
numerically; we will provide a detailed discussion of their VPEs elsewhere \cite{HNM}. 

For $\mu=1$ the particles decouple and the heavier one is subject to the potential 
$v_{22}(x)=-6/{\rm cosh}^2(x)$ which 
is known to have a zero mode and a bound state, the so-called shape mode, at $\omega^2=3$ 
(for $m_2=2$) \cite{Ra82}. This corresponds to the imaginary dependent momentum $k_2=\imu$ 
and to the real independent momentum $k=\sqrt{2}$ since $m_1=1$ and $m_2=2$. This is a 
soliton example of a Feshbach resonance. Hence this state is not explicitly 
counted in the discrete sum in Eq.~(\ref{eq:evac1}). Indeed for this real momentum the Jost
determinant of the positive parity channel is zero and the phase shift
jumps by $\pi$. 

\renewcommand{\arraystretch}{1.8} 
\begin{table}[t!]
	\centering
	\begin{tabular}{c|ccccccccccc}
		\toprule
		$\mu $ & 2.0 & 1.5 & 1.0 & 0.9 & 0.8 & 0.7 & 0.6 & 0.5 & 0.4 & 0.3 & 0.2 
		\\ \colrule
		Eq.~(\ref{eq:sol1})
		& -1.333 & -1.151 & -0.985 & -0.953 & -0.922 & -0.891 & -0.860
		& -0.830 & -0.799 & -0.768 & -0.737 
		\\
		Eq.~(\ref{eq:sol2})
		& --- & ---  & -0.985 & -0.969 & -0.960 & -0.961 & -0.976
		& -1.011 & -1.085 & -1.228 & -1.540
		\\ \botrule
	\end{tabular}
	\caption{\label{tab:Evac}Vacuum polarization energies for the
		Bazeia model as computed from Eq.~(\ref{eq:Evac}).}
\end{table}
\renewcommand{\arraystretch}{1.0}

In Table \ref{tab:Evac} we list the numerical results for the VPEs of the two
soliton solutions. For the simple diagonal case, Eq.~(\ref{eq:vpot1a}), we agree with the 
heat kernel results: the top line in Table \ref{tab:Evac} is obtained by adding the 
$\sigma=2$ and $\sigma=\mu$ entries of Table 4 in Ref. \cite{AlonsoIzquierdo:2012tw}. To 
our knowledge, there are no previous studies on the off-diagonal case, Eq.~(\ref{eq:vpot1b}),
in the literature.

Note that the $\mu=2$ entry corresponds to
$2\times\left(\frac{1}{2\sqrt{3}}-\frac{3}{\pi}\right)\sim-1.3326$,
{\it i.e.}~twice the $\phi^4$--kink vacuum polarization energy for quantum fluctuations
with mass $m_2=2$. The case $\mu=1$ is even more interesting not only because the two solitons 
are identical, but also because it is an application of our general method to the particular 
case of zero off-diagonal potential matrix elements, since $\chi\equiv0$, but
with an actual gap present (see also Appendix \ref{app:decouple}). 
While $v_{11}(x)=-2/{\rm cosh}^2(x)$ is the potential for fluctuations off the
sine-Gordon soliton with mass $m_1=1$, $v_{22}(x)=-6/{\rm cosh}^2(x)$ is the potential for
fluctuations off the $\phi^4$ kink with mass~$m_2=2$. Indeed our numerical result 
equals the sum of these well-established VPEs~\cite{Ra82,Graham:2009zz},
$\left(\frac{1}{2\sqrt{3}}-\frac{3}{\pi}\right)-\frac{1}{\pi}
\sim-0.9846$, providing verification of the threshold treatment in particular, 
since the shape mode of the $\phi^4$ part has disappeared as a manifest bound state. The
coupled channel formalism treats the two uncoupled particles in such a way that the
shape mode of the heavier particle is no longer a zero of the Jost determinant on the 
imaginary $k$-axis. As discussed earlier, this is a kinematical feature induced 
by relating the momenta between the  heavier and lighter particles.

As mentioned, the two solitons are classically degenerate. Obviously including one 
loop quantum corrections favors the second solution. The present conventions yield 
$E_{\rm cl}=\frac{4}{3}$  suggesting that for small enough $\mu$ the soliton, 
Eq.~(\ref{eq:sol2}) would unavoidably be destabilized by quantum corrections. 
These convenient conventions included scaling the fields by $\frac{1}{\sqrt{\lambda}}$, 
where $\lambda$ is the fourth order coupling constant when the coefficient of the 
quadratic order does not contain $\lambda$. However, in quantum field theory the scale 
of the field cannot be chosen freely; rather, it is dictated by the equal-time 
commutation relations. Though this is irrelevant when comparing various VPEs in a given 
model, it must be taken into consideration when combining classical and quantum contributions.
When (re)introducing physical parameters, $E_{\rm cl}$ scales like $\frac{1}{\lambda}$, but 
$E_{\rm vac}$ does not depend on $\lambda$ \cite{Ra82,Graham:2009zz}. It is then clear that 
the occurrence of instability depends on the interaction strength.

Also note that for $\mu=2$ the VPE is twice that of the $\phi^4$ model kink even though it is 
the same soliton configuration. This is due to the addition of $\chi$ type quantum fluctuations.
The case $\mu=2$ is also interesting because there is an additional soliton 
solution
\begin{equation}
\phi^{(3)}=\frac{{\rm sinh}(2x)}{b+{\rm cosh}(2x)}
\qquad {\rm and}\qquad 
\chi^{(3)}=\frac{\sqrt{b^2-1}}{b+{\rm cosh}(2x)}\,
\label{eq:sol3}
\end{equation}
with an arbitrary parameter $b>1$. For $b=1$ this becomes the standard kink 
configuration. The classical energy is the same for all allowed values of $b$. 
As already observed in the heat kernel calculation of Ref.~\cite{AlonsoIzquierdo:2003gh}, 
we find that also the VPE does not vary with $b$. Actually the Jost determinant
itself turns out not to depend on $b$. The reason is simple: for
$\mu=2$ the transformation 
$$
\phi=\frac{\eta_1+\eta_2}{2}
\qquad {\rm and} \qquad 
\chi=\frac{\eta_1-\eta_2}{2}
$$
decouples the potential, Eq.~(\ref{eq:fpot}) into two $\phi^4$ models for $\eta_1$ 
and $\eta_2$ \cite{AlonsoIzquierdo:2003gh}. The configuration of Eq.~(\ref{eq:sol3})
corresponds to individual kinks for $\eta_i$ separated
by ${\rm arcosh}(b)$. The solution $\eta_{1}={\rm tanh}(x)$ and $\eta_{2}(x)=1$ translates 
into $\phi(x)=\left[{\rm tanh}(x)+1\right]/2$ and $\chi(x)=\left[{\rm tanh}(x)-1\right]/2$.  
This case is interesting because it induces the symmetric potential matrix
$$
V(x)=\frac{-3}{{\rm cosh}^2(x)}\begin{pmatrix}
1 & 1 \cr 1 & 1\end{pmatrix}\,,
$$
which has a zero eigenvalue. The vacuum polarization energy must then be computed from 
$F_S$ and $F_A$ defined in Eq.~(\ref{eq:defjostsym}). It is reassuring that our numerical
simulation yields $E_{\rm vac}=-0.6625$, consistent with the known VPE of a single kink.
But then, the case $\mu=2$ is not a stringent litmus test for our approach because it does
not exhibit a mass gap.

As mentioned, for $\mu=\frac{1}{2}$ yet another soliton solution is known 
analytically \cite{Shifman:1997wg}
\begin{equation}
\phi^{(4)}=\frac{(1-a^2){\rm sinh}(x)}{a^2+(1-a^2){\rm cosh}(x)}
\qquad {\rm and} \qquad
\chi^{(4)}=\sqrt{\frac{2}{\mu}}\,\frac{a}{\sqrt{a^2+(1-a^2){\rm cosh}(x)}}
\label{eq:sol4}\end{equation}
with $|a|<1$. The case $a=0$ reproduces the pure kink soliton. Though the case $|a|=1$
solves the field equations, it is not a soliton \cite{Bazeia:1995en} because it is not 
localized. Consequently, the corresponding potential matrix, Eq.~(\ref{eq:vpot}) does not 
vanish at spatial infinity. Comparing the numerical results Tables \ref{tab:Evac} and  
\ref{tab:Evac2}, we see that the limiting case $a\to1$ has an even lower VPE than the 
second soliton, Eq.~(\ref{eq:sol2}), for $\mu=\frac{1}{2}$.

\renewcommand{\arraystretch}{1.8}
\begin{table}[t!]
        \centering
        \begin{tabular}{c|ccccc}
                \toprule
                $a $ & 0.0 & 0.2 & 0.4 & 0.6 & 0.8
                \\ \colrule
                Eq.~(\ref{eq:sol4})
                & -0.830 & -0.841 & -0.878 & -0.949 & -1.089 
                \\ \botrule
        \end{tabular}
        \caption{\label{tab:Evac2}Vacuum polarization energies for the soliton of 
            Eq.~(\ref{eq:sol4}) in the Bazeia model with $\mu=\frac{1}{2}$ as 
            computed from Eq.~(\ref{eq:Evac}).}
\end{table}
\renewcommand{\arraystretch}{1.0}

\section{Conclusion}
\label{sec:concl}

We have generalized the spectral methods for computing the vacuum polarization energy (VPE) 
of static solitons to models containing coupled fields with different mass parameters. This 
result is non-trivial endeavor because the spectral methods rely on the analytic properties 
of scattering data, most prominently the Jost function/determinant $F$, but different masses 
induce a non-analytic relation between the momenta that describe the asymptotic behavior of the 
fields. Taking $k$ to be the momentum associated with the smaller mass, the essential ingredient 
is to establish a relation, Eq.~(\ref{eq:k2}), between the two momenta that fulfills 
$F^\ast(k)=F(-k)$ for all real $k$ and that preserves the analytic properties of $F(k)$ in the 
upper half complex momentum plane, ${\sf Im}(k)\ge0$. We have checked this relation by numerically 
verifying those properties for different potentials. The formulation of the VPE is then similar 
to the case without a mass gap. However, it is indispensable to consider momenta off the real 
axis because the no-tadpole renormalization condition would otherwise induce a singularity for 
models in one space dimension. Once the generalized spectral method is established,
the computation of the VPE is numerically straightforward and does not require any further 
approximation or expansion. We have then applied this formalism to a soliton model with two real 
scalar fields to (i) verify that the approach is consistent and (ii) to show that the 
VPE, {\it i.e.} the quantum correction to the soliton mass, lifts the classical degeneracy.

There are many other soliton models in one space dimension to which the formalism can
be applied. For example, the solitons constructed in Refs.~\cite{Sarkar:1976vr,Montonen:1976yk}
have different classical masses and it would be interesting to see whether 
the respective VPEs obey the same inequality.

Obviously the formalism is not constrained to one space dimension but can be applied in higher
dimensions with cylindrical or spherical symmetry. A prime candidate for future investigation
is the VPE of the 't Hooft-Polyakov monopole \cite{tHooft:1974kcl,Polyakov:1974ek}.

\acknowledgments
H.~W.\ thanks M.~H.~Capraro for discussions during early stages of this project.
H.~W.\ is supported in part by the National Research Foundation of South Africa (NRF) 
by grant~109497. N.~G.\ is supported in part by the National Science Foundation (NSF) 
through grant PHY-1520293.

\appendix

\section{Cauchy integrals of the Jost determinant}
\label{app:cauchy}

\begin{figure}
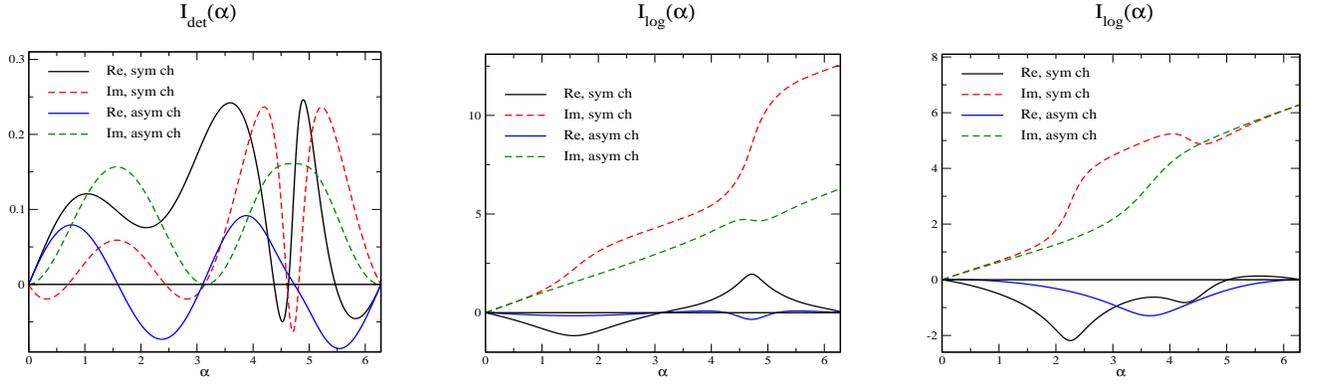

\centerline{
\epsfig{file=cdet.eps,width=5.0cm,height=5cm}\hspace{1.0cm}
\epsfig{file=logdet1.eps,width=5.0cm,height=5cm}\hspace{1.0cm}
\epsfig{file=logdet2.eps,width=5.0cm,height=5cm}}
\caption{(Color online)\label{fig:contour}Contour integrals $I_{\rm det}(\alpha)$
and $I_{\rm log}(\alpha)$ defined in Eq.~(\ref{eq:contint}) for the model 
parameters of Figure \ref{fig:jost2}. The parameters for the line integral,
Eq.~(\ref{eq:contour}) are $k_0=1.2\imu$ and $R=1$ (left and middle panel);
$k_0=0.3+1.4\imu$ and $R=0.6$ (right panel).}
\end{figure}

To further analyze the analytic structure of the Jost determinant we
we take the potential model from Eq.~(\ref{eq:testpot1}) and
consider contours parameterized by a complex center, $k_0$ and a radius $R$:
\begin{equation}
\mathcal{C}(\alpha)\,:\,\, k=k_0+R\,{\rm e}^{\imu\varphi}
\qquad {\rm with}\qquad \varphi=0\ldots\alpha\,,
\label{eq:contour}
\end{equation}
which describe full circles when $\alpha=2\pi$. For these contours we 
numerically compute the line integrals
\begin{align}
I_{\rm det}(\alpha)&=\int_{\mathcal{C}(\alpha)}\, dk\,{\rm det}F(k)
\qquad\qquad {\rm and}\cr
I_{\rm log}(\alpha)&=\int_{\mathcal{C}(\alpha)}\,dk\,
\frac{d}{dk}{\rm ln}\,{\rm det}F
=\int_{\mathcal{C}(\alpha)}\,dk\,
\frac{1}{{\rm det} F(k)}\frac{d}{dk} {\rm det} F(k)\,.
\label{eq:contint}
\end{align}
In Figure \ref{fig:contour} we show typical results for these contour integrals.
For any considered contour (including paths composed of piece-wise linear sections) 
in the upper half complex plane we found results for $I_{\rm det}(\alpha)$ 
as in the left panel. Though there 
are some oscillations as function of $\alpha$, the line integrals vanish when the 
circles are completed.  When the contour enters the lower half plane, the integrals 
cannot be controlled numerically. Also the results from the middle and left panels are 
as expected for a well-behaved Jost determinant. The integral $I_{\rm log}(2\pi)$ 
counts the enclosed zeros in multiples of $2\pi\imu$. The contour of the middle 
figure encloses all bound states identified in Section \ref{sec:num1} and thus we 
find $4\pi\imu$ and $2\pi\imu$ for the symmetric and anti-symmetric channels, 
respectively. The higher bound state in the symmetric channel is not enclosed by the 
contour of the right panel. Hence in that case we obtain $2\pi\imu$ for both channels.

The zeros in the lower half complex momentum plane that correspond to Feshbach resonances
cannot be identified by computing $I_{\rm log}$ because the Jost determinant is not
expected to be an analytic function in that regime. However, we can test whether the zeros 
of Figure \ref{fig:jost_dec} unexpectedly moved to the upper half plane. To this end we
consider a contour defined by the triangle\footnote{A tiny offset into the upper half 
plane improves numerical stability.} $k=1.0+0.0001\,\imu\,\rightarrow\,2.0+0.0001\,\imu\,
\rightarrow\,1.5+\imu\, \rightarrow\,1.0+0.0001\,\imu$ for different values of 
the off-diagonal elements of $V_0$, $v_{12}=0.01$ and $v_{12}=0.2$. In the vicinity of
the Feshbach resonances and the threshold the numerical integration requires tiny step
sizes. In the end the numerical results for $I_{\rm log}$ are indeed compatible with zero.

\section{VPE for decoupled particles}
\label{app:decouple}

In this appendix we will show that $\widetilde{E}_{\rm vac}$,
Eq.~(\ref{eq:evac2}) is additive when the two particles decouple. 
Since Eq.~(\ref{eq:evac2}) is constructed in terms of
the lighter particle's momentum, it suffices to verify that 
\begin{equation}
\widetilde{E}_{\rm vac}=\frac{1}{2}\sum_{\omega_i^2\le m_2^2}
\left[\omega_i-m_2\right]
+\int_0^\infty \frac{dq}{2\pi}\,\left[\sqrt{q^2+m_2^2}-m_2\right]
\frac{d}{dq}\overline{\delta}(q)
\label{eq:app1}\end{equation}
follows from Eq.~(\ref{eq:evac2}) when $V(x)=\begin{pmatrix}0 & 0 \cr
0 & v(x)\end{pmatrix}$ where $\overline{\delta}(q)$ is the phase shift 
for the scattering problem
$$
q^2\psi(x)=-\partial_x^2\psi(x)+v(x)\psi(x)\,,
$$
with $\Delta^{(1)}(q)$, as defined in Eq.~(\ref{eq:born2})
with $V_{11}(x)=0$ and $V_{22}(x)=v(x)$, subtracted. 

Let $n_0$ and $n_1$ be the number\footnote{If 
threshold half-bound states are present $n_0$ and $n_1$ are half-integer and the 
corresponding contributions in the discrete sums are weighted by $\fract{1}{2}$.}
of bound states with $\omega_i^2\le m_1^2$ and $m_1^2<\omega_i^2\le m_2^2$,
respectively. As discussed after Eq.~(\ref{eq:defjostsym}) and in the context of 
Figure \ref{fig:jost_dec}, each bound state with $\omega_i^2\le m_1^2$ adds $\pi$ 
to the phase in $0\le k\le \overline{k}=\sqrt{m_2^2-m_1^2}$ and each bound state with 
$m_1^2<\omega_i^2\le m_2^2$ adds $\pi$ from $k_i=\sqrt{\omega_i^2-m_1^2}$ to $\overline{k}$. 
Hence the phase shift that enters Eq.~(\ref{eq:evac2}) is
\begin{align}
\left(\sum_{s}\delta_s(k)\right)
-\Delta^{(1)}(k)=n_0\pi\left[\theta(k)-\theta(k-\overline{k})\right]
+\pi\hspace{-0.3cm}\sum_{m_1^2<\omega_i^2\le m_2^2}\hspace{-0.3cm}
\left[\theta(k-k_i)-\theta(k-\overline{k})\right]
+\theta(k-\overline{k})\,\overline{\delta}\left(\sqrt{k^2-\overline{k}^2}\right)\,,
\label{eq:app2}\end{align}
where $\theta(k)$ is the Heaviside step function.
The derivative with respect to the momentum $k$ produces various Dirac-$\delta$ functions
that yield discrete contributions in Eq.~(\ref{eq:evac2}). Noting that $\omega(0)=m_1$
and $\omega(\overline{k})=m_2$ we find
\begin{align}
\widetilde{E}_{\rm vac}&=\frac{1}{2}\sum_{\omega_i^2\le m_1^2}
\left(\omega_i-m_1\right)-\frac{n_0}{2}\left(m_2-m_1\right)
+\frac{1}{2}\sum_{m_1^2<\omega_i^2\le m_2^2}\left(\omega_i-m_1\right)
-\frac{n_1}{2}\left(m_2-m_1\right)\cr
&\hspace{1cm}
+\frac{1}{2\pi}\left(m_2-m_1\right)\overline{\delta}(0)
+\int_{\overline{k}}^\infty\frac{dk}{2\pi}\,\left[\sqrt{k^2+m_1^2}-m_1\right]
\frac{d}{dk}\overline{\delta}\left(\sqrt{k^2-\overline{k}^2}\right)\cr
&=\frac{1}{2}\sum_{\omega_i^2\le m_1^2}\left(\omega_i-m_2\right)
+\frac{1}{2}\sum_{m_1^2<\omega_i^2\le m_2^2}\left(\omega_i-m_2\right)
+\int_{\overline{k}}^\infty\frac{dk}{2\pi}\,\left[\sqrt{k^2+m_1^2}-m_2\right]
\frac{d}{dk}\overline{\delta}\left(\sqrt{k^2-\overline{k}^2}\right)\,.
\label{eq:app3}\end{align}
Changing the integration variable to $q$ by
$k=\sqrt{q^2+\overline{k}^2}$ then yields indeed Eq.~(\ref{eq:app1}).

\medskip\noindent
Often the VPE is formulated via integrating Eq.~(\ref{eq:evac2}) by parts \cite{Graham:2009zz}
\begin{equation}
\widetilde{E}_{\rm vac}=\frac{1}{2}\sum_{\omega_i^2\le m_1^2}
\left(\omega_i-m_1\right)
-\int_0^\infty\frac{dk}{2\pi}\,\frac{k}{\omega(k)}\,
\left[\left(\sum_s \delta_s(k)\right)-\Delta^{(1)}(k)\right]
\label{eq:app4}\end{equation}
since this quantity is better accessible numerically. Substituting Eq.~(\ref{eq:app2}) yields
\begin{align}
\widetilde{E}_{\rm vac}&=\frac{1}{2}\sum_{\omega_i^2\le m_1^2}
\left(\omega_i-m_1\right)-\frac{n_0}{2}\omega(k)\Big|_{k=0}^{k=\overline{k}}
-\frac{1}{2}\sum_{m_1^2<\omega_i^2\le m_2^2}\omega(k)\Big|_{k=k_i}^{k=\overline{k}}
-\int_{\overline{k}}^\infty\frac{dk}{2\pi}\,\frac{k}{\sqrt{k^2+m_1^2}}\,
\overline{\delta}\left(\sqrt{k^2-\overline{k}^2}\right)\cr
&=\frac{1}{2}\sum_{\omega_i^2\le m_1^2}\left(\omega_i-m_2\right)
-\frac{1}{2}\sum_{m_1^2<\omega_i^2\le m_2^2}\left(m_2-\omega_i\right)
-\int_0^\infty\frac{dq}{2\pi}\,\frac{q}{\sqrt{q^2+m_2^2}}\,
\overline{\delta}\left(q\right)\,,
\label{eq:app5}\end{align}
which is VPE in the form of Eq.~(\ref{eq:app4}) for a single particle with mass $m_2$.

\end{document}